\documentclass[twocolumn,showpacs,aps,prl,amsfonts]{revtex4}

\usepackage{epsfig}
\usepackage{amsmath}
\usepackage{amssymb}
\usepackage{bm}
\usepackage{graphicx}
\allowdisplaybreaks[1]

\newcommand{\beq}{\begin{equation}}
\newcommand{\eeq}{\end{equation}}
\newcommand{\beqarray}{\begin{eqnarray}}
\newcommand{\eeqarray}{\end{eqnarray}}

\newcommand{\Hc}{\ensuremath{\mbox{H.c.}}} 
\newcommand{\fig}[1]{Fig.~(\ref{#1})} 
\newcommand{\Ref}[1]{Ref.~\onlinecite{#1}} 

\begin{document}

\title{Origin and control of spin currents in a
  magnetic triplet Josephson junction} 
\author{P. M. R. Brydon,${}^1$ Dirk Manske,${}^1$ and M. Sigrist${}^2$} 
\affiliation{${}^1$Max-Planck-Institut f\"{u}r Festk\"{o}rperforschung,
  Heisenbergstr. 1, 70569 Stuttgart, Germany \\
${}^2$Institut f\"{u}r Theoretische Physik, ETH Z\"{u}rich, CH-8093
Z\"{u}rich, Switzerland}

\date{\today}

\begin{abstract}
We study the appearance of a Josephson spin current in a model
triplet superconductor junction with a magnetically-active tunnelling
barrier. We find three distinct mechanisms for producing a spin current, and
we provide a detailed discussion of the symmetry properties and the physical 
origins of each. By combining these three basic mechanisms, we
find that it is possible to exercise fine control over the spin currents. In
particular, we show that unlike the charge current, the spin currents on
either side of the barrier need not be identical.
\end{abstract}

\maketitle

The fabrication and study of devices that combine triplet superconductivity
and magnetism in a controlled manner will undoubtedly lead to new insights
about the unique interplay of these two phases, and possibly also novel
spintronic devices. Much attention has therefore been directed at developing
the theoretical understanding of such
systems~\cite{SingletSOTriplet,TFTprl,Yokoyama,Zhao,Norway,TFTprb}. These 
efforts have already revealed many unconventional effects in the transport
properties of such devices: for example, the Josephson charge current through
a triplet-superconductor--ferromagnet--triplet-superconductor (TFT) junction
depends not only upon the magnitude (as in a singlet junction, see~\Ref{SFS})
but also upon the \emph{orientation} of the ferromagnetic moment at the
barrier~\cite{TFTprl,TFTprb}. The origin of this behaviour is the interaction
of the barrier moment with the spin-structure of the triplet state. Despite
the clear importance of the spin, only a few works have considered the
appearance of a spontaneous spin current in materials where the pairing
has a triplet
component~\cite{Norway,TFTprb,Asano,Rashedi,Vorontsov2008}. 

In this letter we study the appearance of a Josephson spin current in a minimal
model of a TFT junction~\cite{TFTprl,TFTprb}. We find three general mechanisms
for producing a spin current: spin-filtering, misalignment of the triplet
vector order parameters (${\bm{d}}$-vectors), and spin-flipping off a
transverse barrier moment. The symmetry properties of the spin currents due to 
each mechanism are analyzed in detail, and the physical origins are understood
by examining the contributions to the charge and spin currents from each spin
sector. Unlike the Josephson charge current, we show that the Josephson spin
currents on each side of the barrier can be different. By judicious choice of
the tunnelling barrier and the ${\bm{d}}$-vector alignment, we can exercise
fine control over the magnitude, direction and orientation in spin space of
the spin currents on either side of the junction.

\begin{figure}
  \includegraphics[width=0.9\columnwidth]{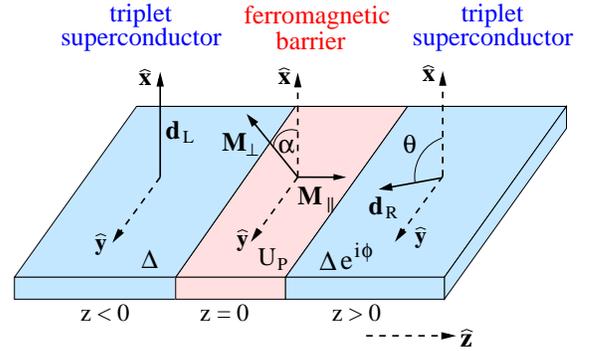}
  \caption{\label{schematic} (color online) Schematic diagram of the model
    junction studied in this work.} 
\end{figure}

The Hamiltonian describing the TFT junction is given by $H=\int
{\cal{H}}(z,z')dz dz'$ with Hamiltonian density
\beqarray
\lefteqn{{\cal{H}}(z,z')} \notag \\
 & = &
\sum_{\sigma}\psi^{\dagger}_{\sigma}(z')\delta(z'-z)\left[-\frac{\hbar^2}{2m}\frac{\partial^2}{\partial{z}^2}-\mu
  + U_{P}(z)\right]\psi^{}_{\sigma}(z) \notag \\
& & +
\left\{\frac{1}{2}\Delta(z,z')\left[e^{-i\theta_\nu}\psi^{\dagger}_{\uparrow}(z')\psi^{\dagger}_{\uparrow}(z)
    -
    e^{i\theta_\nu}\psi^{\dagger}_{\downarrow}(z')\psi^{\dagger}_{\downarrow}(z)\right]
\right. \notag \\
& & \left.+\Hc \right\} -
{\bm{M}}(z,z')\cdot\sum_{\alpha,\beta}\psi^{\dagger}_{\alpha}(z')\hat{\pmb{\sigma}}_{\alpha\beta}\psi^{}_{\beta}(z) \label{eq:Ham}
\eeqarray
where $\nu=L(R)$ for $z', z<0$ ($z',z>0$). $\psi^{\dagger}_{\sigma}(z)$
($\psi^{}_{\sigma}(z)$) is the fermionic creation (annihilation) operator for
a spin-$\sigma$ particle with co-ordinate $z$. The quasiparticles have
effective mass $m$ and chemical potential $\mu$ in the superconductors. The
superconducting gap is antisymmetric with respect to particle interchange, 
i.e. $\Delta(z,z')=-\Delta(z',z)$. As in~\Ref{TFTprl} and~\Ref{TFTprb}, we
assume $p_{z}$-pairing orbitals in both superconductors. The gap obeys a BCS
temperature dependence, and has $T=0$ magnitude $\Delta_{0}$. The condensates
differ by a phase $\phi$. The ${\bm{d}}$-vector on the $\nu$ side of the
junction is parameterized as ${\bm{d}}_\nu=(\cos\theta_\nu,\sin\theta_\nu,0)$
where we take $\theta_{L}=0$ and $\theta_{R}=\theta$ for definiteness
(i.e. the left ${\bm{d}}$-vector defines the $x$-axis). As the
${\bm{d}}$-vectors have no $z$-component, this is the mutual spin orientation
for pairing in the two superconductors, i.e. they are both in an
equal-spin-pairing state with respect to the $z$-axis. At the barrier we have
potential scattering by $U_{P}(z)=U_P\delta(z)$ and magnetic scattering by the
moment
${\bm{M}}(z,z')=(M_{\perp}\cos\alpha,M_{\perp}\sin\alpha,M_\parallel)\delta(z)\delta(z')$;
$\hat{\bm{\sigma}}_{\alpha\beta}$ denotes the vector of Pauli matrices. In
what follows, we set $\hbar=1$ and quote values of $M_\perp$, $M_\parallel$ 
and $U_P$ in units of $m/k_F$ where $k_F$ is the Fermi momentum. A schematic
diagram of the junction is shown in~\fig{schematic}. 

We construct the $4\times4$ Nambu-spin space Green's function
$\check{\cal{G}}(z,z',\Omega;i\omega_n)$ of the junction in terms of
quasiclassical scattering wavefunctions. A full account of this method for
singlet superconductors is given in~\Ref{Kashiwaya2000}; the generalization to
the triplet case is straight-forward. For simplicity, we assume
spatially-constant order parameters within the two superconductors. This
should not qualitatively alter our results. Proceeding from the continuity
equations, we define the Josephson charge ($I^{\nu}_{c}$) and $\mu$-component
of the spin ($I^\nu_{s,\mu}$) currents on the $\nu$-side of the junction 
\beqarray
I^\nu_{c} & = & -i\frac{e}{4{m}}
\frac{1}{\beta}\sum_{n} \int_{\cap} d\Omega
\lim_{z\rightarrow{z'=0}}
\left(\frac{\partial}{\partial{z'}}-\frac{\partial}{\partial{z}}\right)\notag
\\
&& \times \mbox{Tr}\left\{\check{\cal{G}}_{\nu}(z,z',\Omega;i\omega_n)\right\}\\
I^\nu_{s,\mu} & = & i\frac{1}{8{m}}
\frac{1}{\beta}\sum_{n} \int_{\cap} d\Omega
\lim_{z\rightarrow{z'=0}}
\left(\frac{\partial}{\partial{z'}}-\frac{\partial}{\partial{z}}\right) \notag
\\
&& \times \mbox{Tr}\left\{\check{\cal{G}}_{\nu}(z,z',\Omega;i\omega_n)\left|
\begin{array}{cc}
\hat{\sigma}_\mu & \hat{0} \\
\hat{0} & \hat{\sigma}_\mu^{\ast}
\end{array}
\right|\right\}
\eeqarray
where $\int_\cap$ indicates an integration over half the spherical Fermi
surface, $\Omega$ is the solid angle and
$\check{{\cal{G}}}_\nu(z,z',\Omega;i\omega_n)$ is the Green's function for $z, 
z'> (<) 0$ when $\nu=R(L)$. In the absence of chiral symmetry-breaking, the
currents only flow along the $z$-axis.

\begin{figure}
  \includegraphics[width=\columnwidth]{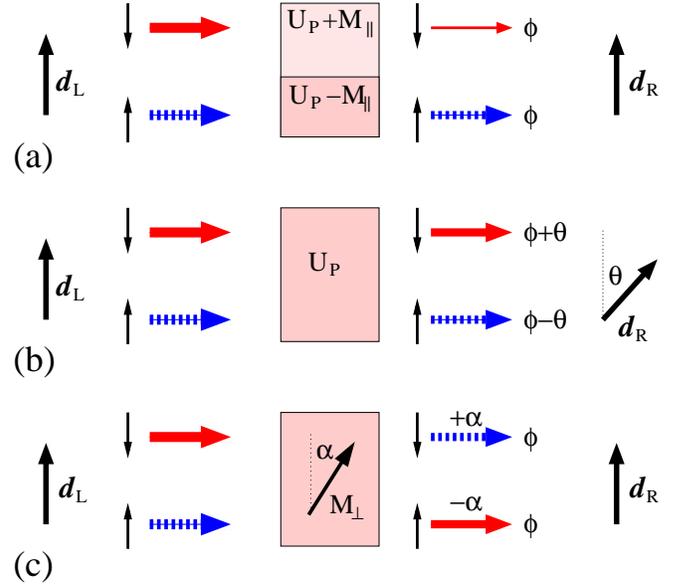}
\caption{\label{cartoon} (color online) Cartoon representation of the three
  basic mechanisms for producing a spin current. (a) Spin-filter: spin-dependent
  effective barrier heights favour transmission through the spin-$\uparrow$
  sector over the spin-$\downarrow$ sector. (b) ${\bm{d}}$-vector
  misalignment: each spin-sector sees a different effective phase difference
  $\phi-\sigma\theta$. (c) Spin-flipping: electron-like quasiparticles
  spin-flipped by a transverse moment acquire a spin-dependent phase factor
  $\sigma\alpha$.} 
\end{figure}

\begin{figure}
  \includegraphics[width=\columnwidth]{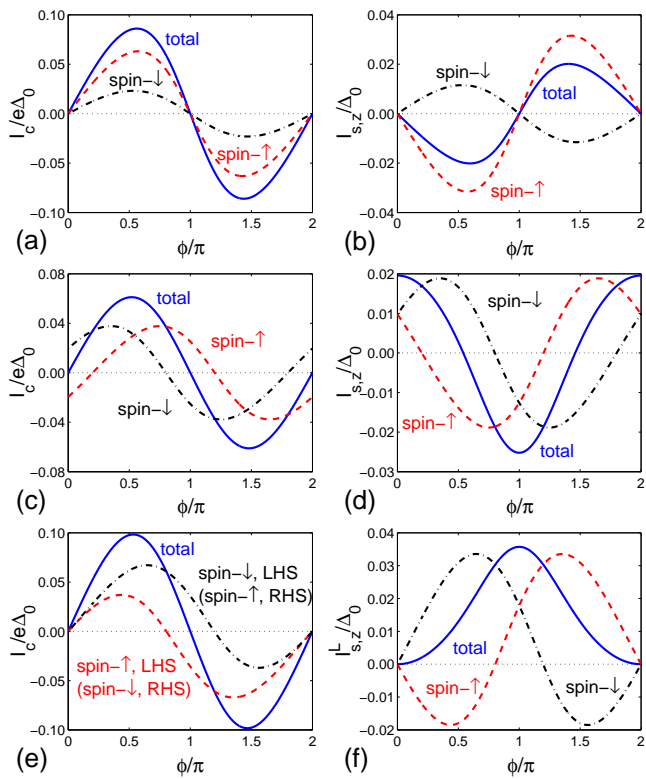}
  \caption{\label{mechanisms} (color online) Current vs phase
    relations for the three basic mechanisms. (a) Charge current and (b) 
    $z$-component of spin current for a spin-filter barrier
    ($M_\parallel=0.5$, $U_P=1$, other parameters vanishing), decomposed
    into contributions from each spin sector. (c) Charge current and (d) spin
    current for a non-magnetic barrier and with misaligned ${\bm{d}}$-vectors
    of the two superconductors ($U_P=1$, $\theta=0.2\pi$, other parameters
    vanishing). (e) Charge current and (f) spin current on the LHS of a
    barrier with a transverse moment ($M_\perp=0.5$, $\alpha=0.3\pi$, other
    parameters vanishing). The spin current on the RHS is identical in
    magnitude but has opposite sign. In all figures we set $T=0.4T_{c}$.}   
\end{figure}

The most direct way to produce a spin current is by
the spin-filtering effect~\cite{TFTprb}. For a spin-filter between
equal-spin-pairing superconductors, we
require both a finite $U_P$ and a component of the magnetic
moment ${\bm{M}}$ in the
spin quantization plane of each superconductor. The component of ${\bm{M}}$
in the spin quantization planes then defines a preferred direction, with the
spin projections along this axis not mixed by the scattering 
at the barrier. Rather, we find spin-dependent effective scattering 
potentials $U_P\pm{|{\bm{M}}|}$. The transparency of the barrier is
therefore different in each spin sector, favouring the transmission of one
spin species of Cooper pair over the other, i.e. the barrier selectively
filters out one spin state, see~\fig{cartoon}(a). By time-reversal symmetry
and mirror 
reflection of the junction in the $x$-$z$ plane, we find that the spin
current is antisymmetric in both $\phi$ and ${\bm{M}}$. We now specialize to
the case when ${\bm{M}}=M_{\parallel}\hat{\bf{z}}$, as this remains a
spin-filter for arbitrary $\theta$. The spin structure of the Josephson
currents are best appreciated by isolating the contributions to the total
charge and spin currents 
from each spin sector: we define the spin-$\sigma$ charge current as  
$I^\nu_{c,\sigma}=(I^{\nu}_{c}-\sigma2e{I^\nu_{s,z}})/2$, and the
spin-$\sigma$ spin current $I^{\nu}_{s,z,\sigma}=-\sigma{I^{\nu}_{c,\sigma}}/2e$. 
For the spin-filter, we find that for $U_P, M_{\parallel}>0$ the magnitude of
$I_{c,\uparrow}$ is greater than $I_{c,\downarrow}$  [\fig{mechanisms}(a)]. As
such, $I^\nu_{s,z,\uparrow}$ and $I^{\nu}_{s,z,\downarrow}$ do not cancel,
giving us the finite spin current shown in~\fig{mechanisms}(b).

Asano has shown that a spin current is also generated in a
non-magnetic junction by misaligning the ${\bm{d}}$-vectors of the
superconductors~\cite{Asano}. This 
produces a gradient of the order parameter in spin space, driving a spin
current in analogy to the situation in superfluid $^{3}$He~\cite{3He}. For
the TFT junction studied here, this results in distinct phase differences in
each of the spin channels,
$\phi_{\sigma}=\phi-\sigma\theta$~\cite{Kwon2004,TFTprb}, as illustrated
schematically in~\fig{cartoon}(b). We show the charge currents through each
spin 
sector in~\fig{mechanisms}(c): they are $2\theta$ out of phase with respect to 
one another but otherwise identical. Although each spin sector has
a finite current flowing through it at $\phi=0$, these cancel and so the total
charge current is vanishing; this is not true of the spin current,
demonstrating that the spin current can flow in the absence of a charge
current [\fig{mechanisms}(d)]. Unlike the spin-filter, the spin current
produced by the ${\bm{d}}$-vector misalignment is symmetric with respect to
$\phi$, which follows from time-reversal symmetry.
When ${\bm{d}}$-vector misalignment is combined
with a spin-filter, the charge currents do
not cancel at $\phi=0, \pi$ and the junction is then in a fractional
state~\cite{TFTprb,Sigrist1995}.

Both the spin-filter and the ${\bm{d}}$-vector misalignment mechanisms
preserve the spin of the tunnelling Cooper pairs. We hence find that, like the
charge current, the spin currents on either side of the junction are the
same. This is not the case, however, for the third mechanism, where the
barrier magnetic moment has a component outside the quantization plane of the
two superconductors. We consider the least complicated such scenario: we have
aligned 
${\bm{d}}$-vectors, and there is only a transverse moment at the junction
(i.e. $M_\perp\neq0$, $M_\parallel=U_P=\theta=0$). We
find that only the $z$-component of the spin current is non-vanishing for
$0<\alpha<\pi/2$. Although the magnitude of 
the spin current is the same on each side of the junction, the \emph{sign} is
reversed, i.e. the $\nu=L$ and $\nu=R$ spin currents flow in opposite
directions. This remarkable result is confirmed by a symmetry  analysis:
treating the spin current as a function only of the angle 
$\alpha$ and the phase $\phi$, we find by time-reversal and inversion symmetry
that $I^{\nu}_{s,z}(\alpha,\phi)=-I^{-\nu}_{s,z}(\alpha,\phi)$~\cite{inversion}. 
For $M_{\perp}=0.5$ and $\alpha=0.3\pi$, we show the charge current
and its decomposition into spin-$\uparrow$ and spin-$\downarrow$ components on
each side of the barrier in~\fig{mechanisms}(e), while the
$z$-component of the spin currents on the LHS are shown
in~\fig{mechanisms}(f). 
The physical interpretation of this behaviour follows by examining the charge
currents through each spin sector:
from~\fig{mechanisms}(e), we observe that the current
through the spin-$\uparrow$ ($\downarrow$) sector on 
the LHS of the junction is equal to the current through the spin-$\downarrow$
($\uparrow$) sector on the RHS. That is, the spin of some of the tunnelling
Cooper pairs is flipped by the transverse moment; this spin-relaxation
  process 
prevents a spin-accumulation occurring at the barrier. 
When a spin-$\sigma$ electron-like (hole-like) quasiparticle tunnelling between
two normal regions 
undergoes a spin-flip at the tunnelling barrier, the transmitted and reflected
waves acquire a phase shift of $+(-)\sigma\alpha$. Although the spin
components of the charge current indicate that the situation is
somewhat more complicated for tunnelling between two superconductors, this
spin-dependent phase shift is ultimately 
the origin of the spin current. This conclusion is immediately apparent from
the presence 
of the normal-normal scattering matrix in the boundary conditions for
the quasiclassical Green's function at a spin-active
interface~\cite{spinactive}. We illustrate the spin-flipping mechanism 
schematically in~\fig{cartoon}(c).

We have now discussed in detail the three basic mechanisms for the production
of a spin current in the TFT junction. We have already seen that the 
sign of the spin current need not be the same on either side of the junction.
By combining the three mechanisms, however, we demonstrate that we can also
vary the magnitude and spin-orientation of the current.

\begin{figure}
  \includegraphics[width=\columnwidth]{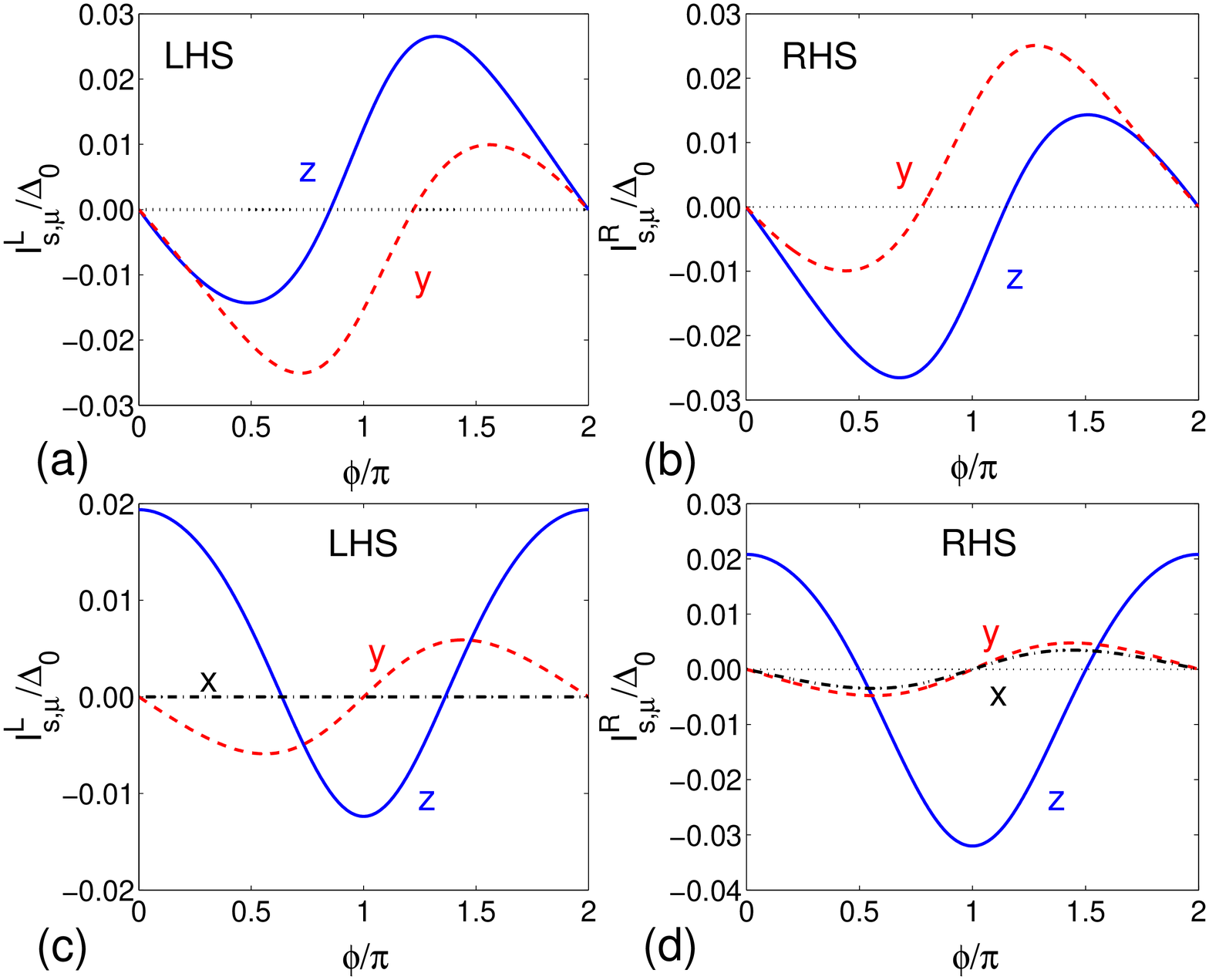}
  \caption{\label{combination} (color online) $\mu$-component of the spin
    currents from 
    combinations of the three mechanisms. Spin currents on the (a) LHS and (b)
    RHS of a junction combining spin-filtering with spin-flipping ($U_P=1$,
    $M_\parallel=M_\perp=0.5$, $\alpha=0.3\pi$, other parameters
    vanishing). Spin currents on the (c) LHS and (d) RHS of a junction
    combining a transverse barrier moment and misaligned ${\bm{d}}$-vectors
    ($M_\perp=0.5$, $\alpha=0.5\pi$, $\theta=0.2\pi$, other parameters
    vanishing). In all figures we set $T=0.4T_{c}$.}  
\end{figure}

Upon combining the transverse moment with a spin-filter ($M_{\perp},
M_{\parallel}, U_P\neq0$, $\theta=0$), we find two significant changes: a
$y$-component of the spin current 
appears in addition to the $z$-component, and the magnitude of the spin
currents on either side of the barrier are no longer the same for
$0<\alpha<\pi/2$.  We show the spin currents on the LHS and RHS of such a 
barrier in~\fig{combination}(a) and (b) respectively.
We consider first the origins of the $y$-component. For $\alpha\neq0,\pi$,
this follows immediately from the spin-filter effect: for aligned
${\bm{d}}$-vectors, the $y$-axis lies within 
the quantization plane of both superconductors. As the magnetization lying in
the $y$-$z$ plane has a finite component along the $y$-axis, the resulting
spin-filter selects a direction with a finite $y$-component, and so
$I^{\nu}_{s,y}\neq0$. This is the only magnetic interaction
at the barrier when $\alpha=\pi/2$, and the spin current is then the same on
either side of the 
junction. For $0<\alpha<\pi/2$, however, there is a finite component of the
magnetization along each axis: the scattering off the components transverse to
the $y$-axis and off those transverse to the $z$-axis also produces a
$y$- and a $z$-component of the spin current respectively. As above, the spin
currents  
produced by the spin-flipping flow in opposite directions on either side of the
junction. The total spin current on each side of the junction is then the
sum of a contribution which is symmetric and a contribution which is
antisymmetric with respect to $\nu$. As the former is antisymmetric
in $\phi$ while the latter is symmetric, and both are $2\pi$-periodic in
the phase difference, we have the relation 
$I^{\nu}_{s,z(y)}(\phi)=-I^{-\nu}_{s,z(y)}(2\pi-\phi)$,
which can clearly be seen in~\fig{combination}.

We finally discuss the combination of a spin non-conserving moment with a
misalignment of the ${\bm{d}}$-vectors. For a finite $\theta\neq0,\pi$, the
quantization planes of the two superconductors only intersect along the
$z$-axis: any barrier moment which has a component transverse to the $z$-axis
is not spin-conserving for the tunnelling Cooper pairs. We therefore consider
the case 
where, in the absence of the ${\bm{d}}$-vector misalignment, the magnetic
scattering produces a $y$-component of the spin current. When the
${\bm{d}}$-vectors are misaligned, a $y$-component of the spin current still
flows on the LHS; on the RHS, however, we have an additional $x$-component of
current, which is related to the $y$-component by
$I^{R}_{s,x}=\tan(\theta)I^{R}_{s,y}$. By using a magnetic moment at the
barrier, we are therefore able to produce a spin current on the RHS with a
component along the axis \emph{perpendicular} to the quantization plane in the
LHS superconductor. 
This situation is displayed in~\fig{combination}(c) and (d), which
respectively show the
spin currents on the LHS and RHS of a spin-filter for the 
$y$-component of spin ($M_\perp, U_P\neq0$,
$\alpha=\pi/2$). Note that the finite $z$-component of the spin
current is composed of contributions due to the misaligned ${\bm{d}}$-vectors
and also from the spin-flip scattering. Unlike the previous cases, there
is no simple 
relationship between the spin currents on either side of the barrier, which is
reflected in~\fig{combination}(c) and (d) by the considerable differences
between the $\nu=L$ and 
$\nu=R$  
$z$-component of the spin current: this is strongly enhanced at $\phi=\pi$ on
the RHS over that on the LHS, while its value at $\phi=0$ is much the same in
the two cases.

In conclusion, we have used the quasiclassical method to study the interaction
of triplet superconductivity and magnetism in a model Josephson junction. Our
main result is the classification of the three distinct mechanisms for
the production of a spin current. The signature properties of each of
these mechanisms are given intuitive physical
explanations by decomposing the total charge
and spin currents into contributions from each spin sector [\fig{cartoon}]. We
demonstrate that in the presence of a spin-relaxing moment at the barrier,
neither the magnitude, direction or spin
orientation of the spin currents on each side of the junction need be
identical. Our work theoretically opens the route for the
inclusion of triplet Josephson junctions in spintronic applications.

The authors thank C. Iniotakis, Y. Asano, K. Bennemann,
B. Kastening, D.K. Morr and F.S. Nogueira for useful discussions. The authors
also thank J. Sirker for his careful reading of the
manuscript. P.M.R.B. gratefully acknowledges the hospitality of A. Simon.


\begin{thebibliography}{99}
\bibitem{inversion}Under mirror reflection in the
  $x$-$y$ plane or inversion, the charge and spin currents on the $\nu$-side
  of the 
  junction are mapped to the $-\nu$-side and their sign is reversed.
\end{thebibliography}
\end{document}